\documentclass[fleqn,usenatbib,useAMS, letters]{mnras}
\usepackage{amssymb,amsmath}
\usepackage{graphicx}
\usepackage{xcolor,natbib}
\usepackage{multirow}
\usepackage[T1]{fontenc}
\usepackage{ae,aecompl}
\usepackage{newtxtext, newtxmath}
\usepackage{float}
\usepackage{subfigure}
\usepackage{longtable}
\usepackage{enumitem}
\usepackage{longtable,listings}
\usepackage[flushleft]{threeparttable}
\usepackage{parcolumns}
\usepackage{bm}
\def\oiii{[O~{\sc iii}]}
\def\nii{[N~{\sc ii}]}

\def\obj{SDSS J2224}

\title[500 pc-scale dual AGN]{SDSS J222428.53+261423.2: unique emission lines properties unveil a sub-kiloparsec dual AGN candidate}

\author[Zheng et al.]
{Qi Zheng$^1$, 
XueGuang Zhang$^{2}$,\thanks{Contact e-mail: \href{mailto:xgzhang@gxu.edu.cn}{xgzhang@gxu.edu.cn}}, 
QiRong Yuan$^1$,\thanks{Contact e-mail: \href{mailto:yuanqirong@njnu.edu.cn}{yuanqirong@njnu.edu.cn}},
Paola Severgnini$^3$,
Cristian Vignali$^{4,5}$\\
$^1$School of Physics and Technology, Nanjing Normal University, No. 1,
        Wenyuan Road, Nanjing, 210023, P. R. China \\
$^{2}$School of Physical Science and Technology, Guangxi University,
          No. 100, Daxue East Road, Nanning, 530004, P. R. China \\
$^{3}$INAF-Brera Astronomical Observatory, Via Brera 28, 20121 Milano, Italy   \\
$^{4}$Dipartimento di Fisica e Astronomia Augusto Righi, Alma Mater Studiorum, Università degli Studi di Bologna, Via Gobetti 93/2, I-40129 Bologna, Italy  \\  
$^{5}$INAF-Osservatorio di Astrofisica e Scienza dello Spazio di Bologna, Via Gobetti 93/3, I-40129 Bologna, Italy
  }

\begin{document}

\label{firstpage}
\pagerange{\pageref{firstpage}--\pageref{lastpage}}

\maketitle

\begin{abstract}
In this paper, we presented a detailed analysis of the Sloan Digital Sky Survey optical spectrum of a new sub-kpc scale dual AGN candidate SDSS J222428.53+261423.2
(=SDSS J2224). The target is one of the few AGNs with all the optical narrow emission lines characterized by double-peaked profiles and with peak separations in velocity units of about 930 km/s.
If the double-peaked narrow emission lines (DPNELs) are due to a dual AGN in \obj, the estimated physical separation between the two cores is about 500 pc.
Meanwhile, three alternative explanations are also discussed in this paper, however, we can not find solid evidence to completely rule them out.
Our results support the presence of a sub-kpc dual AGN with DPNELs in all lines, indicating a key episode of galaxy merging evolution at sub-kpc scale.
\end{abstract}

\begin{keywords}
galaxies: active - (galaxies:) quasars: emission lines -galaxies: nuclei - galaxies: individual
\end{keywords}

\section{Introduction}
Merging of galaxies is an essential process during the formation and evolution of galaxies \citep{Be80, Si98, Ma19, 
Ru21, De23, Zh23}, leading to the possible formation of dual Active Galactic Nuclei (AGNs) with relative separations of thousands to hundreds of parsecs down to the still widely discussed supermassive black hole binary systems on scale of sub-pc \citep{Gr15Na,De19,Se20,Ko21,zxga22,De23}. 
When the spatial separation is large enough (kpc-scale), the dual AGN with two independent narrow line regions (NLRs) can produce double-peaked profiles in the optical spectrum. 

There are well-discussed individual objects with double-peaked narrow emission lines (DPNELs) in the literature. 		
\citet{Zh04} firstly showed a dual AGN in a type 2 quasar, SDSS~J1048+0055, with a combination of 
double-peaked [O~{\sc iii}] and two discrete radio sources. Although systematic searches of objects with DPNELs have been carried out so far to find dual AGNs \citep{Ge07,Xu09,Mc11,Wa09,Co12,Ba12,Ba13,Li13,Wo14}, the efficiency in finding dual AGN systems in DPNELs is still controversial \citep{Sh11a, Fu12, An13, Co18, Ru19, Du22}. DPNELs can indeed be produced also by outflows, jets or rotating disks, associated to a single AGN.

\begin{figure}
	\centering\includegraphics[width=8cm,height=5cm]{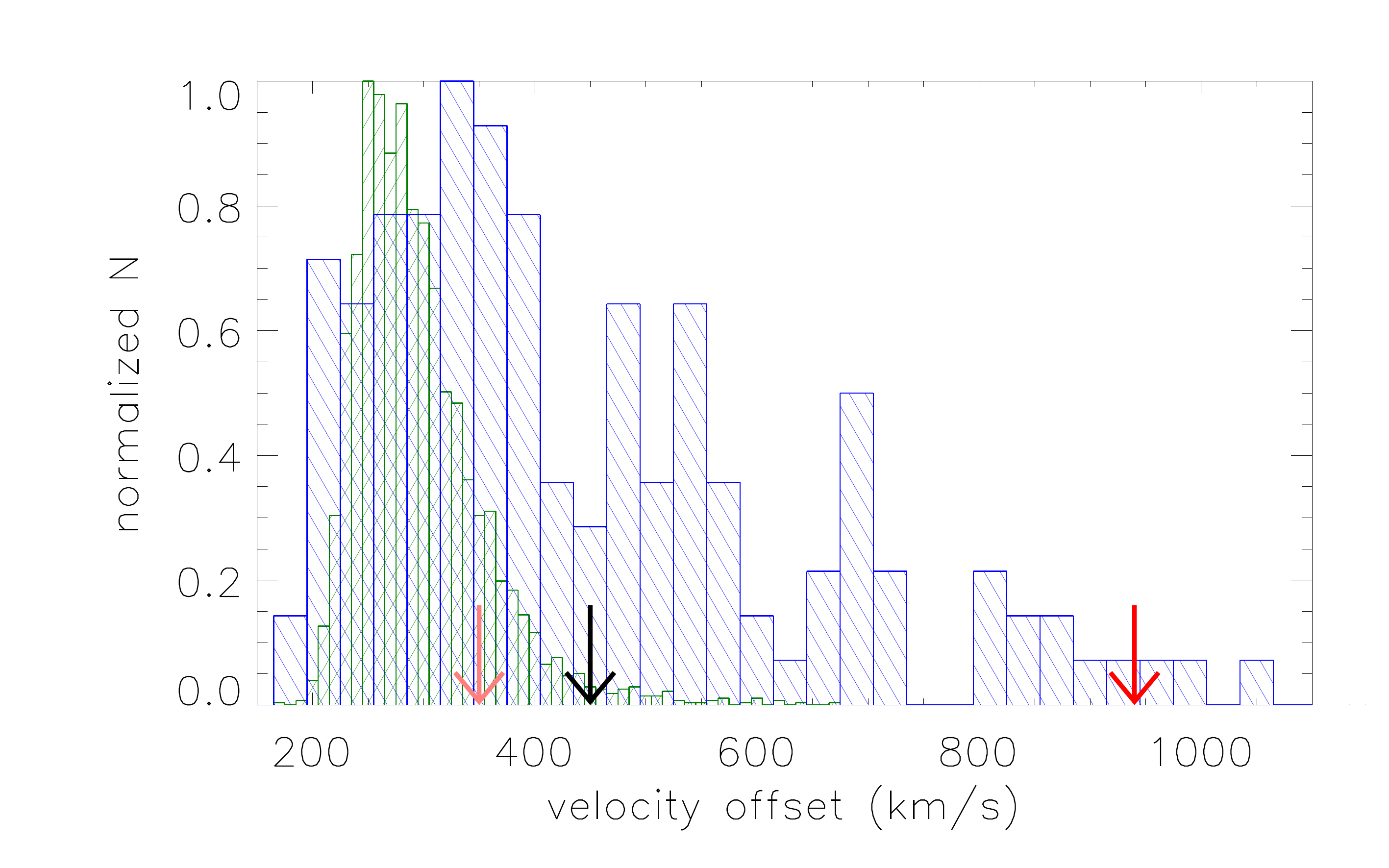}
	\caption{The distributions of $\Delta\upsilon$ of DPNELs. The histograms filled with green and blue lines represent the distributions of \citet{Ge12} sample and \citet{Sm10} sample, respectively. The red arrow marks the peak separation of SDSS J2224, while the black and pink arrows mark the peak separations for the object presented by \citet{Xu09} and by \citet{Se21}, respectively.	
	}
	\label{fig1}
\end{figure}

Among the sample of objects with DPNELs, \citet{Ge12} have reported so-far the largest sample of objects from SDSS DR7 (Sloan Digital Sky Survey, data release 7), including 81 type 1 AGNs, 837 type 2 AGNs and 2112 non-active galaxies.
Figure~\ref{fig1} shows the distributions of peak separations between the blue-shifted and red-shifted components ($\Delta\upsilon$) for the DPNELs in the \citet{Ge12} sample (green lines) with mean $\Delta\upsilon$ $\sim$300 km/s 
($\Delta\upsilon$<700 km/s). 

The $\Delta\upsilon$ distributions for the sample of type 1 AGNs with DPNELs by \citet{Sm10} (blue lines) are also shown in Figure~\ref{fig1}. 
It is clear that there are several type 1 AGNs having DPNELs with $\Delta\upsilon$ around 1000 km/s.
Assuming that DPNELs result from a dual AGN, a larger $\Delta\upsilon$ traces a smaller spatial separation between the two AGNs.
It is interesting to note from Figure~\ref{fig1} that there are no sources having $\Delta\upsilon$ larger than 1500 km/s. Such high velocities correspond to relative separations smaller than the typical NLR inner size (hundreds of parsecs) \citep{Lgl13,apjs22,De22}; in this case, in the presence of dual AGN, the two NLRs are already merged and the source appears as a single NLR AGN. Following this argument, some of the DPNELs populating the high tail of the $\Delta\upsilon$ distributions in Figure~\ref{fig1} could host sub-kpc dual AGNs in the pre-merger NLR phase, i.e. the most direct progenitor of binary AGNs \citep{De19}.

As discussed in \citet{Xu09}, the presence of double-peaked profiles in all narrow emission lines is considered as a strong indication of the presence of dual AGNs. 
SDSS J131642.90+175332.5 \citep{Xu09} is the first type 2 AGN showing double-peaked features in all optical narrow emission lines with $\Delta\upsilon$ of about 400$\sim$500 km/s, leading to unique spectroscopic characteristic of at least three apparent peaks around [S~{\sc ii}] doublet. Similarly, SDSS J143132.84+435807.20 (type 2 AGN) is another promising sub-kpc scale dual AGN candidate showing double-peaked profiles in all optical narrow emission lines (relative separation lower than 500 pc, \citealt{Se21}). Many other objects show double-peaked profiles in all narrow emission lines, as have been observed so far, but all of them exhibit   $\Delta\upsilon$ around 300$\sim$400 km/s (e.g. Figure 1 of~ \citealt{Wa09}).

Here, we presented the case of SDSS J222428.53+261423.2 (=SDSS
J2224). This source is a type 1 AGN with all optical narrow emission lines showing double-peaked
features and populating the high tail of the velocity distributions showed in Figure~\ref{fig1} with a $\Delta\upsilon$ value about two times larger than the objects quoted above. These properties make this target a very promising sub-kpc dual AGN in the late stage before the binary phase. The analysis and relevant results are presented in Section 2, while possible physical models for SDSS~J2224 are discussed in Section 3.
All the objects from the \citet{Sm10} sample with $\Delta\upsilon$> 750 km/s have been also analysed and the results are discussed in Section 3, and their additional information on the detailed emission line properties is provided in the Appendix.
Summary and conclusions are reported in Section 4. In this paper, we have adopted the cosmological parameters of $H_{0}=70 {\rm km/s/Mpc}$, $\Omega_{\Lambda}=0.7$ and $\Omega_{\rm m}=0.3$.

\section{Data Analysis}

\begin{figure*}
\centering\includegraphics[width=18cm,height=12cm]{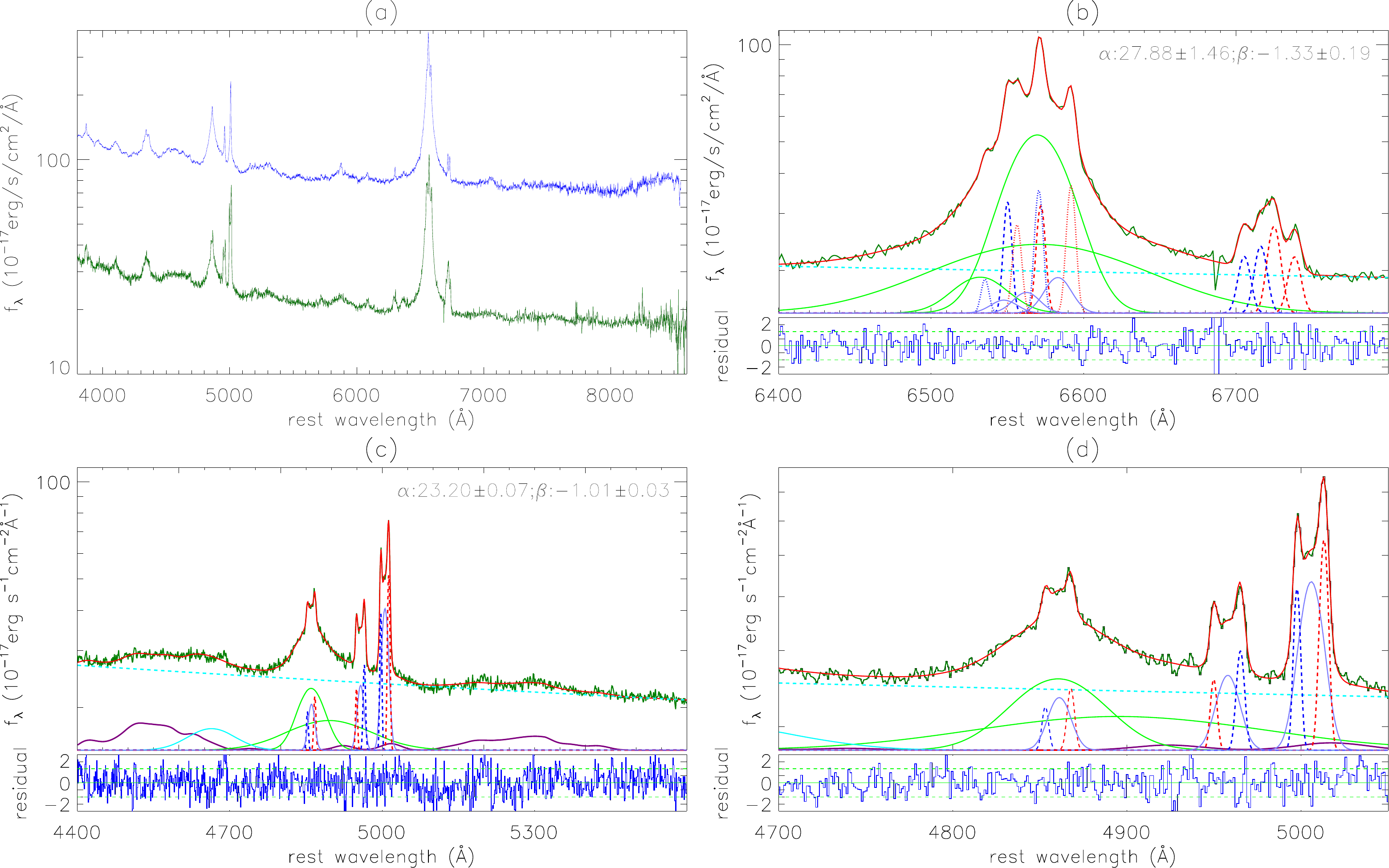}
\caption{Panel (a): rest-frame optical spectrum of SDSS J2224 (green line) compared with the composite spectrum of SDSS quasars by \citet{Va01} (blue line).
Panel (b): rest-frame spectrum of SDSS J2224 from
6400\AA~to 6800\AA.	
Panel (c): rest-frame spectrum of SDSS J2224 from
4400\AA~to 5600\AA, including Fe~{\sc ii} emission lines.
Panel (d): rest-frame spectrum of SDSS J2224 from
4700\AA~to 5050\AA~in order to more clearly show the 
double-peaked profiles of H$\beta$ and \oiii~emission lines.
In the top of panel (b), (c), (d), the solid dark green lines represent the line spectra; the solid red lines represent the best fitting results; the dashed cyan lines represent the continuum emissions; the dashed blue and red lines represent the blue-shifted and red-shifted components of narrow H$\alpha$, H$\beta$, \oiii~and [S{\sc ii}] doublets;
the dotted blue and red lines represent the blue-shifted and red-shifted components of [N{\sc ii}] doublet;
the solid blue lines show the extended components; the solid green lines show broad H$\alpha$ and H$\beta$ emission lines; 
the solid purple and cyan lines represent the Fe~{\sc ii} and He~{\sc ii} emission lines, respectively.
The parameters $\alpha$ and $\beta$ of the power law $\alpha\times(\frac{\lambda}{5100\textsc{\AA}})^{\beta}$ are listed in panel (b) and (c), respectively.
In the bottom of panel (b), (c), (d), the solid blue lines show the residuals calculated by the line spectra minus the best fitting results and then divided by the uncertainties of SDSS spectra; the solid and dashed green lines represent 0 and $\pm$1, respectively.
}
\label{fig2}
\end{figure*}

SDSS J2224 is a type 1 AGN at $z\sim0.206$ (the SDSS pipeline determined value). The spectrum (Plate-MJD-Fiberid: 7648-57329-0836) is similar to the composite spectrum of SDSS quasars derived by \citet{Va01} (see panel (a) of Figure \ref{fig2}) with both broad H$\alpha$ and H$\beta$ emission lines.
We mainly focused on the most prominent emission lines, H$\beta$, \oiii, H$\alpha$, \nii~and [S{\sc ii}] with the rest wavelength ranges from
4400 to 5600 \AA~and from 6400 to 6800 \AA.
The following model function is applied to describe the emission lines.


For continuum emission underneath H$\beta$ (H$\alpha$) emission lines, a single power law component is utilized. 
For each narrow emission line, two narrow Gaussian functions (the second moment $\sigma$<400 km/s) plus one extra Gaussian function ($\sigma$<800 km/s) are applied 
to describe double-peaked profiles and the probably extended component as shown in \citet{Gr05,Sh11b}.
For each broad Balmer emission line, three\footnote{After completing the fitting procedure for describing the broad H$\beta$, it is determined that only two components are reliable, and the line flux of the third broad component is smaller than its corresponding uncertainty. Thus, two broad Gaussian functions suffice to describe the broad H$\beta$.} broad Gaussian functions ($\sigma$>800 km/s) are applied.
For optical Fe~{\sc ii} features, the shifted, broadened and strengthened Fe~{\sc ii} template proposed by \citet{kp10} is applied (refer to \citet{Zh22} for more details). 
For He~{\sc ii} emission line, a single Gaussian function is applied.

For the double-peaked features in each forbidden line doublet and each narrow Balmer line, the corresponding shifted components share the same line widths in velocity space. 
Owing to the complicated line profiles around H$\alpha$, the corresponding shifted components in the narrow H$\alpha$ and [N{\sc ii}] doublet are constrained to have the same line widths in velocity space.	 
The offset central wavelengths and line widths of the extended components are the same in velocity space in all narrow emission lines.

The emission lines of \obj~around H$\beta$ and H$\alpha$ can be well measured based on the Levenberg-Marquardt least-squares minimization technique (the known $\sc{MPFIT}$ package), the fitting results are shown in Figure \ref{fig2}  ($\chi^2/dof$=0.77).
The line parameters, along with their uncertainties, expressed as 1$\sigma$ errors calculated from the covariance matrix using $\sc{MPFIT}$, are reported in Table 1.
The peak separations, along with their corresponding uncertainties, are determined based on central wavelengths and their respective uncertainties, and are provided in Table 1.

Similar model functions and constraints are utilized to characterize the emission lines of the 13 objects in \citet{Sm10} sample using a similar fitting procedure, as illustrated in the Appendix.

In addition to the model function discussed above, a new model function is applied to describe the core components 
by using a single narrow Gaussian component for each narrow emission line ($\chi^2/dof$ is 2.02). Employing the F-test technique, we conclusively determined, with a confidence level exceeding 5$\sigma$, that two narrow Gaussian functions are necessary for an accurate representation of the double-peaked profiles, as opposed to a single Gaussian function.


The average $\Delta\upsilon$ value for our target is $\sim$930 km/s, marked as a red arrow in Figure \ref{fig1}.
SDSS J2224 not only shows double-peaked profiles in all the observed narrow emission lines, supporting the presence of an AGN pair, but the high $\Delta\upsilon$ between the red-shifted and blue-shifted emission line components makes it a good candidate to be sub-kpc scale dual AGN.

\begin{table}
\caption{Line parameters of main emission lines}
\begin{center}
\begin{tabular}{ccccc}
\hline\hline
\multicolumn{5}{|c}{narrow emission lines}\\ \hline
lines   & $\lambda_0$ & $\sigma$ & flux & $\Delta\upsilon$ \\ \hline
\multirow{2}{*}{H$\beta$}      & 4853.2$\pm$0.3   & 111.0$\pm$8.6 & 22.0$\pm$4.9 & \multirow{2}{*}{900.7$\pm$37.0}  \\
{ }     & 4867.8$\pm$0.3   & 123.4$\pm$6.2 & 32.5$\pm$6.5 & { }  \\
\hline
\multirow{2}{*}{[O\sc{iii}]$\lambda$5007\AA}    &4997.7$\pm$0.1  & 107.8$\pm$6.0 & 115.1$\pm$7.2  & \multirow{2}{*}{927.8$\pm$12.0} \\
			{ }     & 5013.2$\pm$0.1  & 131.8$\pm$6.0 & 200.5$\pm$11.1  & {  } \\
			\hline
\multirow{2}{*}{H$\alpha$}      & 6550.2$\pm$0.2  &110.0 & 110.9$\pm$10.7 & \multirow{2}{*}{995.6$\pm$36.6} \\
			{ }     & 6572.0$\pm$0.6  &123.4 & 115.6$\pm$11.0 & { } \\
			\hline
\multirow{2}{*}{[N\sc{ii}]$\lambda$6584\AA}     & 6570.5$\pm$0.5 & 110.0 & $\ast$129.2  & \multirow{2}{*}{977.4$\pm$31.9} \\
			{ }     & 6591.9$\pm$0.2 & 123.4 & 148.2$\pm$10.1 & { } \\
			\hline
\multirow{2}{*}{[S\sc{ii}]$\lambda$6717\AA}     & 6705.6$\pm$0.3  & 174.1$\pm$13.4 & 70.7$\pm$4.8 & \multirow{2}{*}{870.8$\pm$35.7} \\
			{ }     & 6725.1$\pm$0.5  & 183.1$\pm$8.9 & 129.2$\pm$12.1 & { } \\
			\hline
\multirow{2}{*}{[S\sc{ii}]$\lambda$6731\AA}     & 6716.3$\pm$0.7  & 174.1 & 87.2$\pm$10.9 & \multirow{2}{*}{993.7$\pm$44.6} \\
			{ }     & 6738.6$\pm$0.3  & 183.1 & 73.9$\pm$ 3.7 & { } \\  \hline
\multicolumn{5}{|c}{broad emission lines}\\ \hline
lines & $\lambda_0$ & \multicolumn{2}{|c}{$\sigma$} & flux  \\ \hline
\multirow{2}{*}{H$\beta$}   & 4860.4$\pm$0.9  & \multicolumn{2}{|c}{1604.0$\pm$80.2}  & 532.9$\pm$34.8 \\ 
{} & 4897.5$\pm$5.3 & \multicolumn{2}{|c}{4843.0 $\pm$265.3} & 684.2$\pm$57.5 \\ \hline
\multirow{3}{*}{H$\alpha$} & 6532.1$\pm$6.0  & \multicolumn{2}{|c}{822.6$\pm$23.4} &   189.0$\pm$78.7 \\ 
{} & 6570.6$\pm$1.6 & \multicolumn{2}{|c}{2961.3$\pm$137.1} & 1513.3$\pm$66.0    \\ 
{} & 6569.8$\pm$1.3 & \multicolumn{2}{|c}{950.6$\pm$45.7} & 1975.8$\pm$191.2  \\ \hline
Fe\sc{ii} & $\star$0.0$\pm$ 0.0  & \multicolumn{2}{|c}{1156.1$\pm$ 117.1} & 896.6$\pm$102.8 \\ \hline
He\sc{ii} & 4664.2$\pm$3.7  & \multicolumn{2}{|c}{2905.9$\pm$230.4}  & 278.1$\pm$20.9   \\ \hline
\multicolumn{5}{|c}{extend components in narrow emission lines}\\ \hline
lines & $\lambda_0$ & \multicolumn{2}{|c}{$\sigma$} & flux  \\ \hline
H$\beta$ & 4861.3  & \multicolumn{2}{|c}{400.3} & 90.2$\pm$13.0  \\ \hline
[O\sc{iii}]$\lambda$5007\AA &  5006.8$\pm$1.2 & \multicolumn{2}{|c}{400.3$\pm$10.8} & 419.9$\pm$15.9  \\ \hline
H$\alpha$ & 6563.1  & \multicolumn{2}{|c}{400.3} & 50.0$\pm$10.9   \\ \hline
[N\sc{ii}]$\lambda$6584\AA & 6583.8  & \multicolumn{2}{|c}{400.3} & 91.3$\pm$60.0  \\ \hline
\hline

\end{tabular}
\begin{tablenotes}
\item
[*]The line flux with $\ast$ has great uncertainties due to superposition of peak of H$\alpha$ and [N{\textsc{\sc ii}}] doublet.
$\lambda_0$ is the central wavelength in units of \AA, $\sigma$ is the second moment of narrow line in units of km/s,
flux is the line flux in units of $10^{-17}{\rm erg/s/cm^{2}}$, and $\Delta\upsilon$ is the peak separation in units of km/s.
The central wavelength with $\star$ shows the shifted velocity to be zero for the optical Fe{\sc ii} emission lines.
\end{tablenotes}
\end{center}
\end{table}

To derive the physical separation between the two putative AGN cores, assuming a circular motion, the velocity offset $\Delta\upsilon$ related to orbital motion can be described as:
\begin{equation}
\begin{split}	
	\Delta\upsilon
	=(\sqrt{\frac{G(m_{1}+m_{2})}{D}})\times\sin(i)\times\cos(\phi),
\end{split}
\end{equation}
with $m_1$ and $m_2$ as total mass of each core, $D$
as the intrinsic spatial distance between two cores, $i$ as the inclination angle of the rotation plane, and $\phi$ as the orientation angle. 
Although determining the $i$ and $\phi$ can be challenging, the upper limit of $D$ can be assessed by considering the maximum value of $sin(i)\times cos(\phi)=1$. 
Once the total masses $m_1$ and $m_2$ are provided, the upper limit of the distance can be estimated. 
The $m_1$ and $m_2$ can be estimated by two methods.

The total stellar mass $M_{\ast}$ can be determined through the $M_{\ast}$-$\sigma_{\ast}$ relation, where $\sigma_{\ast}$ represents the stellar velocity dispersion \citep{Lw13,Za16,Da22}. However, it is noteworthy that this $M_{\ast}$-$\sigma_{\ast}$ relation in the literature is primarily established for quiescent galaxies, raising uncertainties about potential biases when applied to AGNs. To address this, we constructed a dedicated sample of AGNs to test the $M_{\ast}$-$\sigma_{\ast}$ relation; additional information about this sample is available in the Appendix. Since SDSS J2224 is a prototypical type 1 AGN, the SDSS spectrum exhibits no discernible spectroscopic features from the host galaxy. 
Moreover, the broader component in the \oiii~doublet is commonly associated with outflows from AGN central regions \citep{Ta01,Gr05,Zh21}.
In such cases, the most viable option is to employ the line width of narrow emission line as a proxy for the stellar velocity dispersion \citep{Ne96,Gr05}. 
It is worth noting that there is often a flat dependence of $\sigma_{\ast}$ on radius, as extensively discussed and demonstrated in studies such as \citet{Tr02,Gr05,Be15}. This suggests that the line width obtained from the SDSS fiber spectrum can reasonably serve as an estimate for $\sigma_{\ast}$ around the effective radius. The estimated $M_{\ast}$ values based on the line widths of \oiii~ are $10^{10.59\pm 0.30}\rm M_\odot$ and $10^{10.77\pm 0.28}\rm M_\odot$, with uncertainties derived from the full widths at half maximum of the $M_{\ast}$ distributions, as illustrated in Figure 2 in the Appendix. Accepting these $M_{\ast}$ values as $m_1$ and $m_2$, the upper limit for the distance between the two AGNs is 491$\pm$365 pc. The uncertainties encompass both the uncertainties associated with $M_{\ast}$ and the separation between the peaks.

Besides, $m_1$ and $m_2$ can be estimated based on the central black hole (BH) mass.
Since the broad line region contribution associated with each of the two putative nuclei is not distinguishable, it is hard to estimate the virial BH mass for the AGNs with broad emission lines \citep{Gr051}.
Recent results by \citet{Fe00,Ge00,Di05,Jo09,Sa19,Be21} have 
demonstrated that also dual cores follow the $M_{\rm BH}-\sigma_{\ast}$ relation. We used this 
relation in \citet{Ko13} and the widths of the narrow emission lines as tracers of $\sigma_{\ast}$ finding $10^{7.44\pm 0.22}\rm M_\odot$ and $10^{7.76\pm 0.17}\rm M_\odot$ for the two putative BH masses.
The uncertainties take into account the uncertainties on both the \oiii~line widths and the equation in \citet{Ko13}.	
The bulge mass can be estimated by the relation between the stellar mass of the bulge $M_{\rm \star,bul}$ and the BH mass reported by \citet{Sc19}.
In the case of SDSS J2224, we found $\sim 10^{10.22\pm0.26}{M_\odot}$ and $\sim10^{10.45\pm0.21}{M_\odot}$ for the two bulges.
The uncertainties come from both uncertainties of BH masses and parameters in equation in \citet{Sc19}.
Considering $M_{\rm \star,bul}$ as $m_1$ and $m_2$ in equation (1), the upper limit of $D$ is 211$\pm$110 pc.

According to the $D$ estimated by two methods, the two host galaxies, including their bulges, might be almost completely merged. Consequently, the estimated values for $m_1$ and $m_2$ should be considered as upper limit values in determining the upper limit for D.

We estimated the continuum luminosity at 5100 \AA~to be $(1.45\pm0.01)\times10^{44}\rm erg/s$ of SDSS J2224 based on the continuum emission shown in panel (c) of Figure \ref{fig2}. By comparing this value with the total \oiii~emission line luminosity $(1.20\pm0.06)\times10^{42}\rm erg/s$, we found a continuum-to-\oiii~ ratio of 121$\pm$6 with uncertainty determined by those of the continuum emission and \oiii~emission line luminosity.
This value is three times lower with respect to the typical 
continuum-to-\oiii~ratio 457$\pm$40 of normal quasars from \citet{Sh11b}. The smaller ratio in SDSS J2224 could be explained by 
assuming that one of the two putative AGNs is strongly obscured in the optical band, i.e., type 2 AGN.


\section{Discussions}
As discussed in \citet{Xu09}, an object with double-peaked profiles in all of its narrow emission lines is a particularly good candidate of dual AGN.
Different physical scenarios have been invoked in the literature to explain the presence of DPNELs in AGN, e.g., a rotating disk-like NLR, chance superposition and AGN-driven outflow \citep{Ba13}.
Below, we would discuss if these alternative scenarios can explain the observational properties of SDSS J2224.

According to the model of rotating disk, the double-peaked profiles can be observed as blue-shifted and red-shifted components when the gas in NLRs moving toward and away from us \citep{Liu10}.
Rotational eﬀects due to a NLR associated with a single SMBH are difficult to be detected
in face-on type 1 AGN; to this end, intermediate viewing angles are needed.
An inherent correlation between line width, velocity offset, and flux is expected to emerge, driven by influence of gravitational potential on the motion of gas within the NLR \citep{Xu09,Sm12}.
For a homogeneous disk, because of the Doppler effect, the blue-shifted components should not be weaker than the red-shifted components, which is opposite to the fact in \oiii~in \obj. 
However, an inhomogeneous disk can probably accounted for this result.
Further research is needed to ascertain the homogeneity or inhomogeneity of the disk in \obj.

In the superposition hypothesis \citep{Xu09}, these are two unrelated AGNs that are by chance seen along the same line of sight \citep{Bo09,De14}. 
However, the superposition scenario is not the preferred explanation for SDSS J2224 for the following reason.
The foreground object would obscure the background one, resulting in a more significant extinction in the background one. 
The Balmer decrement regarding the broad components is 3.0$\pm$0.5, so the broad line region is not heavily obscured by dust.
The Balmer decrement (flux ratio of narrow H$\alpha$ to narrow H$\beta$ emission lines) is about 3.56$\pm$1.09 in the red-shifted component, while is 5.04$\pm$2.01 in the blue-shifted component, and the significance level is smaller than $10^{-14}$ for similar Balmer decrement confirmed by student's t-statistic technique.
So it seems to disfavor the superposition scenario.

For the AGN-driven outflow model, a powerful outflow is essential to exert a pronounced impact on NLRs, particularly concerning low-ionization lines. The combined influences of shock driven forces, thermal, magnetic and radiation result in a stratified velocity structure \citep{La21,Ko22}. If the DPNELs of SDSS J2224 arise from different regions, variations in both peak separations and line widths would naturally arise. Nevertheless, the velocities and line widths of different narrow emission lines, as listed in Table 1, do not exhibit significant difference.
In the case of jet-induced outflow, the fluxes of blue-shifted components should be stronger than red-shifted components because of the boost effect. 
The flux of red-shifted component is rather stronger in \oiii.
It seems to disfavour the AGN-driven outflow hypothesis in basic and symmetric scenarios.
While the current findings can not rule out the asymmetric case, such as asymmetric bipolar outflow.


In addition to SDSS J2224, there are 13 more objects with DPNELs exhibiting peak separations larger than 750 km/s in the sample from \citet{Sm10}. However, an explicit explanation for the origin of the double peaks is not provided in their study.
We re-fitted the SDSS spectra of these 13 objects, and showed
the spectra along with the adopted models in the Appendix. 
A brief discussion on their emission line properties is given below. 
Our findings suggest that none of them can be unequivocally considered as robust dual AGN candidates, for four main reasons: 
$\bm{(1)}$Two of these objects (Plate-Mjd-Fiberid: 1687-53260-0428, 2947-54533-0050) have large flux ratios between the red-shifted and blue-shifted components in \oiii.
Based on the F-test technique, confidence level is smaller than 3$\sigma$ to support double-peaked features in \oiii~in the objects.
$\bm{(2)}$ The velocity offset of the narrow H$\beta$ emission line in each of the three objects (Plate-Mjd-Fiberid: 0581-52356-0575, 0885-52379-0449, 0980-52431-0419) is 100 km/s larger than both two narrow components of \oiii, which disfavors the dual AGN model.
In the case of dual AGNs, narrow H$\beta$ emission line should have the similar velocity offset as one of the \oiii~narrow components.
$\bm{(3)}$ The H$\beta$ absorption line in the host galaxy of each of the three objects (Plate-Mjd-Fiberid: 1716-53827-0140, 2019-53430-0219, 10438-58142-0151) aligns with both the narrow H$\beta$ and one narrow component of \oiii, displaying a velocity offset near zero based on the spectroscopic redshift.
This favours AGN-driven outflow model.
$\bm{(4)}$ The narrow H$\beta$ emission line and one of the double-peaked profiles in \oiii~
of each of the five objects (Plate-Mjd-Fiberid: 0607-52368-0625, 1426-52993-0110, 1711-53535-0153, 2225-53729-0002, 9586-57787-0272) have similar velocity offsets. If they are treated as dual AGNs, there should be one core with one set of narrow emission lines and the other one with only \oiii. 
The AGNs with only \oiii, which have never been reported previously, are very special.
In addition, although several of the 13 objects in \citet{Sm10} sample could appear as promising dual AGNs, the double-peaked features are not related to the dual cores, which will be discussed in our future work.

The physical separation of about 500 pc between the two AGN cores in SDSS J2224 indicates that the merger is in the dynamic friction stage and probably triggers black hole activity.
To search for traces of a merger in the host galaxy of our target, we visually inspected the DESI (the Dark Energy Spectroscopic Instrument, \citealt{Dey19}) image in the r-band (Figure 3 in the Appendix). It exhibits a possibly faint tidal tail, providing a potential indication of the ongoing merging process. 
It is impractical to obtain spectroscopic features of narrow emission lines caused by orbital motion due to the 3.3 Myr orbital motion period and tens of Myr
timescale of galaxy merging from kpc to pc scale as simulated in \citet{Pf17,So22}.
If feasible, a future sample of double-peaked narrow emission line objects with significant peak separations at various redshifts could offer valuable statistical insights.

In the literature, there are numerous candidates of kpc-scale dual AGNs (or dual galaxies) and sub-pc binary black holes. However, confirmed sub-kpc scale dual AGN systems, which are expected outcomes of intermediate evolutionary stages in the galaxy merging process, remain exceedingly rare. 
Future research efforts aimed at identifying additional sub-kpc scale dual AGN candidates are crucial for gaining detailed insights into the underlying factors contributing to this rarity. Estimating the real fraction of sub-kpc dual AGNs holds significance in testing models related to galaxy evolution and mergers. These systems are the more direct precursors to binary black holes, which are challenging to detect, providing robust evidence of gravitational waves.


\section{Summary and Conclusions}
In this paper, we analysed the optical spectroscopic properties of \obj, a quasar at z=0.206. The target exhibits  double-peaked profiles in all of its narrow emissions and
very large peak separations for all narrow emission lines. Main summary and conclusions are as follows.
\begin{itemize}
\item Following the results found and discussed by \citet{Xu09,Se21}, the properties of DPNELs in all lines could be considered as good tracers of the presence of a dual AGN. 
\obj~is a type 1 AGN with DPNELs in all lines. It exhibits the largest peak separation
($\sim$930 km/s) among reported good candidates of dual AGN systems with DPNELs in all lines.
\item Assuming the presence of a dual AGN, 
the total stellar mass $M_{\ast}$ of the two cores in \obj~can be estimated as $10^{10.59\pm0.30}{\rm M_\odot}$ and $10^{10.77\pm0.28}{\rm M_\odot}$ through the $M_{\ast}$-$\sigma_{\ast}$ relation. 
Based on the $M_{\ast}$ and the peak separation of \obj, the upper limit to the distance between two cores results to be $\sim 500$ pc in \obj.
\item 
The stronger line flux and smaller Balmer decrement in the red-shifted component compared to the blue-shifted component, along with the similarity in peak separations and line widths in different DPNELs, seem to disfavor the basic and symmetric scenarios of the rotating disk model, superposition model, and AGN-driven outflow model. However, the current results can not exclude the possibilities of asymmetric cases.
\item We reanalysed the optical spectra of the objects in the \citet{Sm10} sample showing double-peaked profiles only in \oiii~and peak separations beyond 750 km/s. We found that these objects are not good candidates to be dual AGN systems due to large line flux ratio of red-shifted and blue-shifted components in double-peaked profiles, no velocity offset in absorption and emission lines, and different velocity offset between \oiii~and narrow H$\beta$.
\item Recognized as a robust candidate for dual AGNs within sub-kpc distances, SDSS J2224, with its substantial peak separation, may represent the most direct precursors of binary AGNs. If SDSS J2224 is confirmed as a dual AGN system, it would witness a crucial episode in the evolution of galaxy mergers on sub-kpc scales.
\end{itemize}

\section*{Acknowledgements}
We gratefully acknowledge the anonymous referee for giving us constructive comments and suggestions to greatly
improve our paper.
We do appreciate great suggestions from professor Komossa, S. at Max-Planck-Institut für Radioastronomie and professor Tavecchio, F. at INAF/Brera Astronomical Observatory to improve the paper.
Zheng wants to express her gratitude for the three months spent at Brera Astronomical Observatory and for the hospitality of italians.
This work is supported by the National Natural Science Foundation of China (Nos. 12273013, 12173020, 12373014) and  INAF 2022 large grant dual and binary supermassive black hole in the multi-messenger era: from galaxy mergers to gravitational waves and from the agreements ASI-INAF n.2017-14-H.0.
We acknowledge financial contribution from the Postgraduate Research \& Practice innovation Program of Jiangsu Province (Grant No.KYCX23\_1678).
We have made use of the data from SDSS DR16.
The SDSS DR16 website is (\url{http://skyserver.sdss.org/dr16/en/home.aspx}).
The MPFIT website is (\url{http://cow.physics.wisc.edu/~craigm/idl/idl.html}).

\section*{Data Availability}
The data underlying this article will be shared on reasonable request to the corresponding authors
(\href{mailto:xgzhang@gxu.edu.cn}{xgzhang@gxu.edu.cn}).

\section{appendix}
\subsection{Appendix A}
The SDSS spectra of the 13 AGNs by \citet{Sm10} with peak separations beyond 750 km/s are shown in the left panels of Figure \ref{fig11}. 
For 6 objects showing dominant host galaxy contributions, we fitted the continuum emission by using one power law function plus
simple stellar population (SSP) models by \citet{Br03} including 
13 population ages (from 5 Myr to 12 Gyr) and 3 metallicities (Z=0.008, 0.05, 0.02).
The best fitting results are shown with solid red lines in the left panels of Figure \ref{fig11}. 
For AGN dominant objects, the continuum is fitted by a simple 
power law model.
After subtracting the starlight component (if present), we fitted the double-peaked \oiii~emission lines with two narrow Gaussian functions and one extended Gaussian function (the middle panels of Figure \ref{fig11}).  
In the right panels of Figure \ref{fig11}, the best fitting emission line models are re-plotted in the velocity space. The positions related to the velocity offsets of absorption lines (if present) are also marked. 


\begin{figure*} \centering\includegraphics[width = 18cm,height=19.2cm]{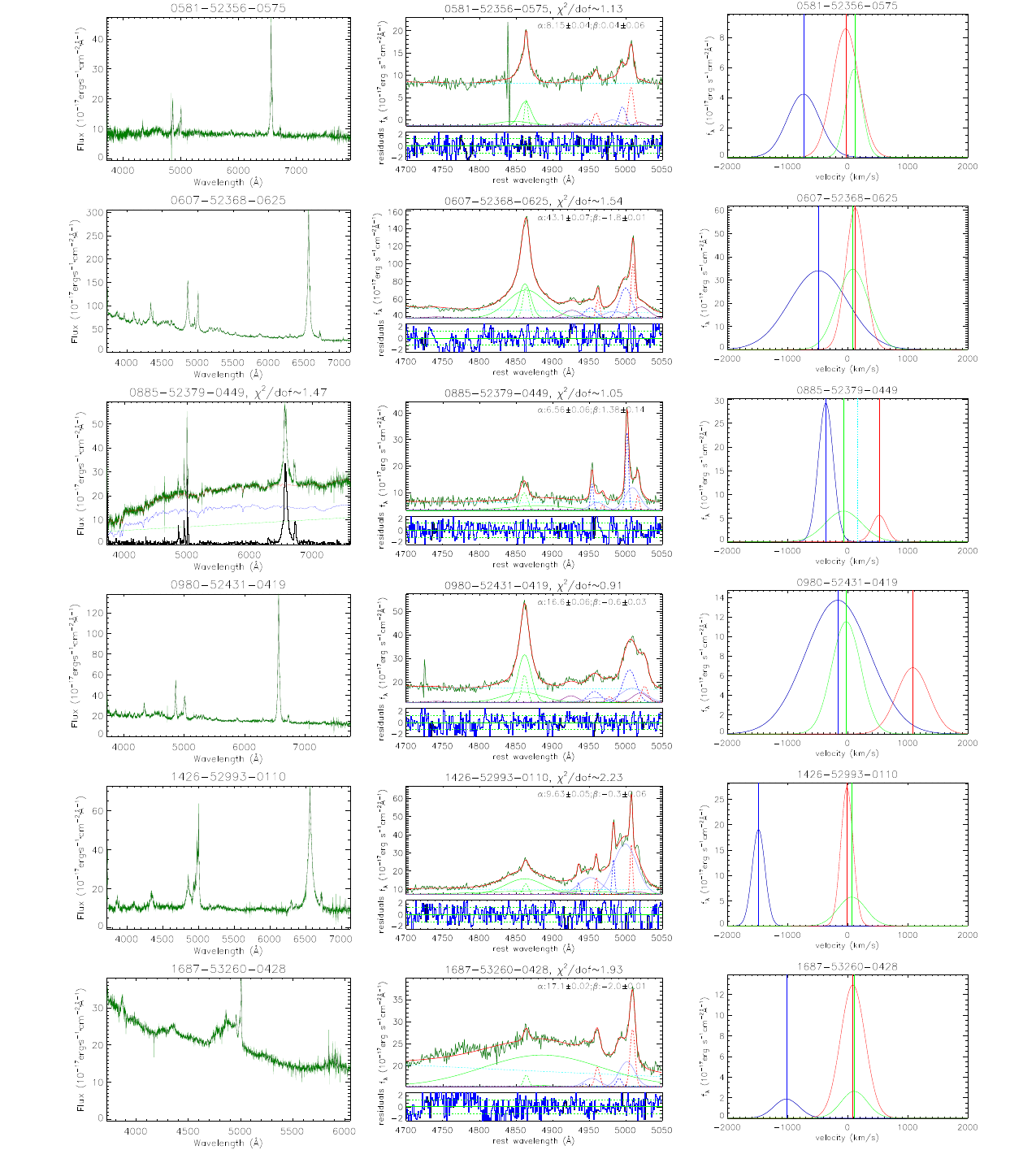} \caption{Left panels: spectra of the 13 AGNs with peak separations larger than 750 km/s reported in \citet{Sm10}.
		Middle panels: rest-frame spectra around the
		H$\beta$+\oiii~emission lines.  
		Right panels: properties of H$\beta$+\oiii~emission lines and/or H$\beta$ absorption lines in velocity space.
		In the left panels, the solid dark green lines show the SDSS spectra with Plate-MJD-Fiberid listed in title, 
		the solid red lines indicate the the best fitting results  (corresponding $\chi^2$/dof listed in title), the solid blue lines show starlights (if present), the solid green lines show AGN continuums (if present), and the solid black lines show the line spectra after subtracting AGN continuums and starlights (if present).
		In the top of the middle panels, the solid dark green lines represent the line spectra after subtracting starlights (if present),
		the solid red lines show the best fitting results (corresponding $\chi^2$/dof listed in title), the dashed cyan lines represent continuum components, the solid green lines show broad H$\beta$, the dashed green lines show narrow H$\beta$, the solid purple lines show Fe{\sc ii} emission lines, the solid blue lines represent extended \oiii~components, and the dashed blue and red lines show the blue-shifted and red-shifted components of \oiii, respectively.  
		The parameters $\alpha$ and $\beta$ of the power law $\alpha\times(\frac{\lambda}{5100\textsc{\AA}})^{\beta}$ are listed in the top of the middle panels.
		In the bottom of the middle panels, the solid blue lines show the residuals calculated by the line spectra minus the best fitting results and then divided by uncertainties of SDSS spectra, and the horizontal solid and dashed green lines show residuals=0,±1, respectively.
		In the right panels, the solid green lines represent narrow H$\beta$ emission lines, the dashed cyan lines show the velocity offsets of absorption lines (if present), and the solid blue and red lines represent the blue-shifted and red-shifted components of \oiii~in velocity space, respectively.
	} \label{fig11} \end{figure*} \setcounter{figure}{2}
\begin{figure*} \centering\includegraphics[width = 18cm,height=24cm]{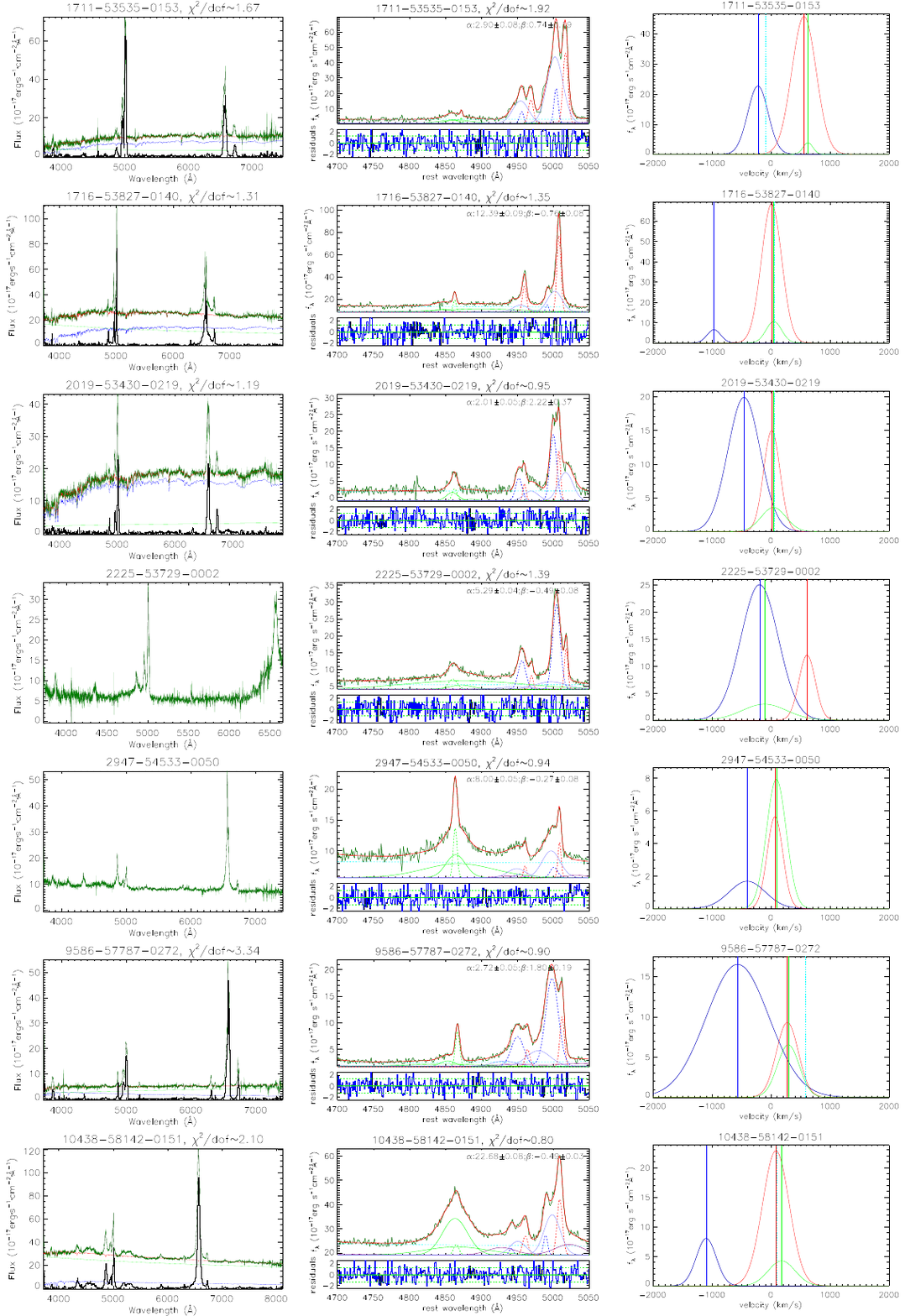} \caption{-- to be continued} \label{fig11} \end{figure*} 
\label{lastpage}

\subsection{Appendix B}
A sample of 12974 narrow emission line AGNs are collected using the SDSS SQL tool based on three criteria: redshift $z < 0.35$, a signal-to-noise ratio (S/N) greater than 5, and fluxes of narrow H$\alpha$, H$\beta$, [O~{\sc iii}], [S~{\sc ii}] emission lines exceeding five times their uncertainties. The narrow emission line parameters and $\sigma_{\ast}$ are obtained from the SpecObjAll database, while the total stellar mass $M_{\ast}$ is retrieved from the databases of stellarMassStarformingPort, stellarMassPCAWiscBC03, and stellarMassPCAWiscM11 databases.

\begin{figure*}
	\centering\includegraphics[width=16cm,height=5cm]{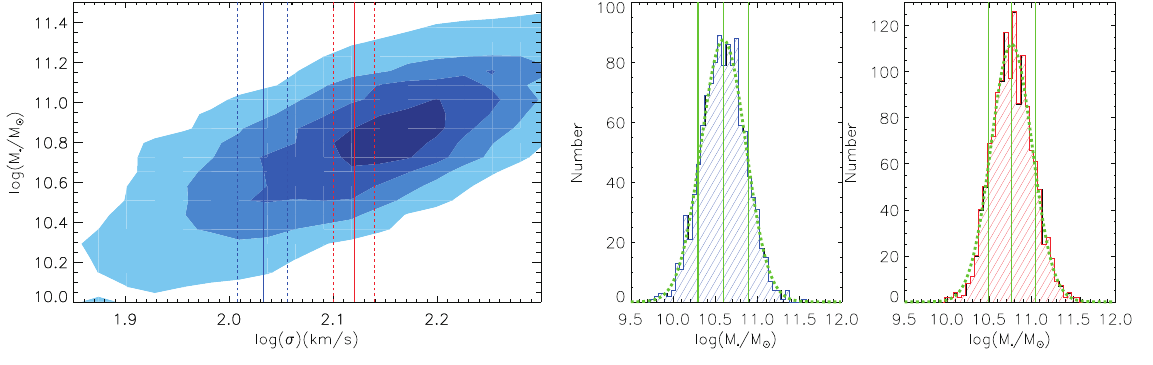}
	\caption{Left panel: the $M_{\ast}$-$\sigma_{\ast}$ relation of 10616 AGNs from SDSS DR16. 
		Middle panel and right panel: the distributions of $M_{\ast}$ determined by line widths of blue-shifted component (the middle panel) and red-shifted component (the right panel).	
		In the left panel, the four contour levels show the corresponding 0.3, 0.5, 0.7 and 0.9 of the two-dimensional volume, and the solid and dashed blue (red) lines represent the line width of the blue-shifted (red-shifted) component and the corresponding uncertainty of [O~{\sc iii}] for SDSS J2224.  
		In the middle and right panels, the dashed green lines mark the best Gaussian fitting of $M_{\ast}$ distributions for SDSS J2224, and the vertical solid green lines mark the mean values and the full widths at half maximum of $M_{\ast}$ distributions. 	
	}
	\label{fig12}
\end{figure*}

To minimize the $M_{\ast}$ bias across different databases, we only considered AGNs with a mass deviation from various databases less than 13\% of the average $M_{\ast}$. Subsequently, 10616 AGNs are selected for our final sample. Their $M_{\ast}$-$\sigma_{\ast}$ relation is reported in the left panel of Figure \ref{fig12}. The distributions of $M_{\ast}$-$\sigma_{\ast}$ within the line widths plus/minus their corresponding uncertainties of the red-shifted and blue-shifted components in [O~{\sc iii}] are presented in the middle and right panels of Figure \ref{fig12}. The mean values and the full widths at half maximum of the $M_{\ast}$ distributions are accepted as the total stellar masses and their corresponding uncertainties for the dual AGN.

\subsection{Appendix C}
\begin{figure}
	\centering\includegraphics[width=8cm,height=8cm]{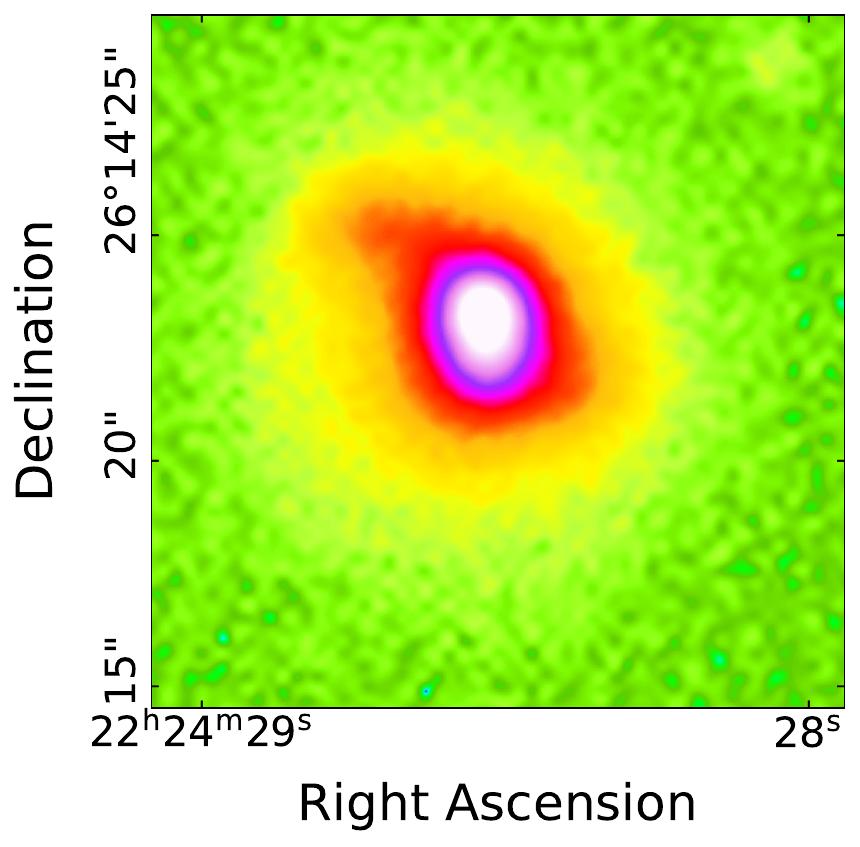}
	\caption{Image of SDSS J2224 in the r-band from the Dark
		Energy Spectroscopic Instrument (DESI) legacy imaging surveys \citep{Dey19}. 
	}
	\label{fig13}
\end{figure}

We collected the r-band photometric image from the Dark Energy Spectroscopic Instrument (DESI) legacy imaging surveys, shown in Figure \ref{fig13}.

\end{document}